\documentclass[a4,twocolumn,aps,pra,floatfix,showpacs,noshowkeys,epsfig,graphics]{revtex4}%
\usepackage{graphicx}
\usepackage{amsmath}
\usepackage{amsfonts}
\usepackage{amssymb}
\usepackage{bm}
\newcommand{\ket}[1]{|{#1} \rangle}
\newcommand{\bra}[1]{\langle {#1}|}
\newcommand{\braket}[2]{\langle {#1}|{#2}\rangle}
\newcommand{\ad}{a^\dagger}
\begin{document}
\title{Entanglement between qubits induced by a common environment with a gap}
\author{Sangchul Oh}\email{scoh@kias.re.kr}
\author{Jaewan Kim}\email{jaewan@kias.re.kr}
\affiliation{School of Computational Sciences, 
             Korea Institute for Advanced Study, Seoul 130-722, Korea}
\date{\today}
\begin{abstract}
We study a system of two qubits interacting with a common environment, 
described by a two-spin boson model. We demonstrate two competing roles of 
the environment: inducing entanglement between the two qubits and making 
them decoherent. For the environment of a single harmonic oscillator, 
if its frequency is commensurate with the induced two-qubit coupling strength, 
the two qubits could be maximally entangled and the environment could be 
separable. In the case of the environment of a bosonic bath, the gap of 
its spectral density function is essential to generate entanglement between 
two qubits at equilibrium and for it to be used as a quantum data bus.
\end{abstract}
\maketitle

%% Introduction

Entangled states, showing nonlocal quantum correlations between subsystems, 
are indispensable in quantum information processing~\cite{Nielsen00}. 
A two-qubit interaction is needed to implement two-qubit gates and 
to make entangled states. In general, there are two types of two-qubit 
couplings: direct coupling such as dipole coupling in NMR or Coulomb 
coupling in superconducting charge qubits, and indirect coupling mediated 
by another quantum system like a harmonic oscillator~\cite{Makhlin01}, 
a cavity mode~\cite{Zheng00}, or even an electron gas~\cite{Craig04}. 
A mediating system is expected to be  separable from the qubits, so that 
its dynamics can be ignored. Recently, Cubitt {\it et al.} showed two qubits 
could be entangled by continuous interaction with a mediating particle in 
a separable state~\cite{cubitt03}.

Entanglement of qubits and an environment causes the decoherence of qubits,
one of the biggest obstacles in quantum information processing. It was 
believed that qubits should be isolated from the environment. However, 
it is shown that two qubits without a direct coupling but interacting with
the environment could be entangled~\cite{Braun02} and even in the steady 
state~\cite{Benatti03}. Once qubits become entangled in the steady state, 
they will never be disentangled due to the irreversible process
described in Ref.~\cite{Benatti03}. This entangling process can not be used 
for the two-qubit coupling scheme to implement two-qubit gates. 

\begin{figure}
\includegraphics[scale=0.7,angle=0]{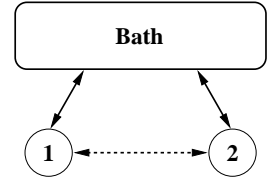}
\caption{Two circles refer to two qubits and an oval is an environment.
         The coupling between qubits and the common environment (solid arrow) 
         induces the indirect interaction between two qubits (dashed arrow).}
\label{Fig:two-spin-boson}
\end{figure}

Like a single mediating particle, could an environment comprised of 
many particles induce the two-qubit coupling? Could it work as a quantum 
data bus to implement two-qubit gates at equilibrium? In this paper, 
we address this problem by studying a two-spin boson model shown in 
Fig.~\ref{Fig:two-spin-boson}. The spin-boson model has been extensively 
studied to describe dissipative two-state systems~\cite{Leggett87,Weiss98}. 
Here we consider two types of the environment: a single mode case and 
a multi mode case. 

\paragraph*{\it Single mode case.---}Consider a system of two qubits 
interacting with a single harmonic oscillator with frequency $\omega$, 
described by the Hamiltonian~\cite{cubitt03}
\begin{align}
H = \hbar\omega\ad a + \hbar\lambda(\sigma_{1z} + \sigma_{2z})(\ad + a)\,,
\end{align}
where $\ad$ is a creation operator of the harmonic oscillator, $\sigma_{iz}$ 
are the $z$-component of the Pauli spin matrices of the {\it i}-th qubit, 
and $\lambda$ the coupling constant. 
The Hamiltonian is solved through 
the canonical transformation $e^X = \exp[\, \frac{\lambda}{\omega}
(\sigma_{1z} + \sigma_{2z})(\ad - a)\,]$~\cite{Leggett87,Mahan90}. 
The transformed Hamiltonian reads
\begin{align}
\label{Eq:Trans_Hamil}
\widetilde{H} 
= e^X H e^{-X}
= \hbar\omega\ad a - I_\text{\rm eff}\,\sigma_{1z}\sigma_{2z} - I_{\rm eff}\,,
\end{align}
where $I_{\rm eff} \equiv 2\hbar\lambda^2/\omega$ is the indirect two qubit 
coupling induced by the harmonic oscillator. A total state is also transformed 
as $\ket{\widetilde{\Psi}} = e^X \ket{\Psi}$.

The time evolution of the total state is given by 
$\ket{\Psi(t)} = e^{-X}\,e^{-i\widetilde{H}t/\hbar}\,e^{X}\,\ket{\Psi(0)}$. 
Let us suppose an initial state is given by a product state
$\ket{\Psi(0)} = \ket{\psi(0)}\otimes\ket{0}_E$
where $\ket{\psi(0)} = a\ket{00} + b \ket{01} + c\ket{10}+ d\ket{11}$ 
is the initial state of the two qubits and $\ket{0}_E$ the ground state of 
the harmonic oscillator. Then the total state at time $t$ is given by
\begin{align}
\ket{\Psi(t)} 
&= e^{i\theta t - i\Gamma_I(t)}\,\Bigl[\, 
     a\ket{00}\otimes\ket{\alpha(t)}_E + d\ket{11}\otimes\ket{-\alpha(t)}_E 
              \nonumber\\[8pt]
&+ e^{-i2\theta t + i\Gamma_I(t)}\, 
    (b\ket{01} + c\ket{10})\otimes\ket{0}_E\, \Bigr]\,,
\label{Eq:psi_t}
\end{align}
where the phase $\theta\equiv I_{\rm eff}/\hbar = 2\lambda^2/\omega$ is due to 
the indirect coupling between the two qubits, the phase 
$\Gamma_I(t)\equiv\left(\frac{2\lambda}{\omega}\right)^2 \sin\omega t$ is 
due to the action of a displacement operator on the coherent state, and 
$\ket{\alpha(t)}_E$ is a coherent state with 
$\alpha(t) =\frac{2\lambda}{\omega}(e ^{-i\omega t} -1)$. 
Notice that the subspace spanned by $\{\ket{01},\ket{10}\}$ forms a decoherence 
free subspace~\cite{Duan97}. In general, Eq.~(\ref{Eq:psi_t}) shows 
entanglement among the two qubits and the harmonic oscillator, depending on 
the parameters: $\omega$ and $\lambda$. 

\begin{figure}[htbp]
\includegraphics[scale=1.0]{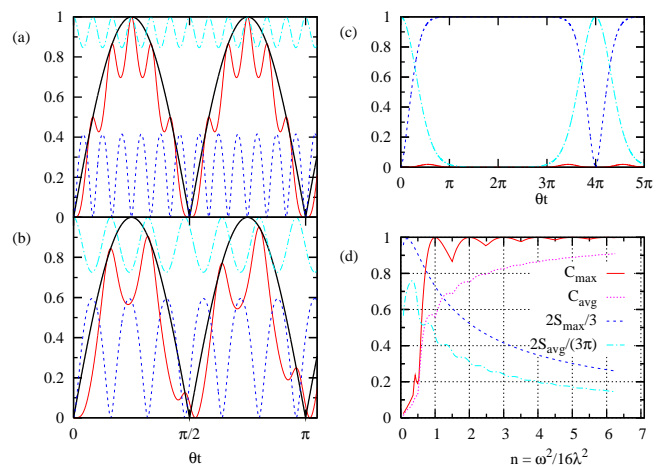}
\caption{(color online). Concurrence $C$ (red solid), $C_{\rm ideal}$ 
         (black thick solid), entropy $2S/3$ (blue dashed), and the overlap 
         $\exp[-\Gamma_R(t)]$ (cyan dashed-dotted) as a function of time for 
         $\omega/\lambda = 4\sqrt{3}$ (a), $5$ (b), and $1$ (c). 
         (d) Maximum concurrence $C_{\rm max}$, maximum entropy $S_{\rm max}$, 
         average concurrence $C_{\rm avg}$, and average entropy $S_{\rm avg}$ 
         as a function of $n=(\omega/4\lambda)^2$ for $0\le \theta t < \pi/2$.}
\label{Fig2}
\end{figure}

For $\omega\gg\lambda$, $\ket{\alpha(t)}$ is very close to $\ket{0}$. This 
means the overlap $\braket{0}{\alpha(t)} = \exp[-\Gamma_{R}(t)] \approx 1$ 
where $\Gamma_R(t) \equiv (\frac{2\lambda}{\omega})^2(1-\cos\omega t)$ 
goes to 0. Also the phase $\Gamma_I(t)$ becomes zero. It is interesting 
that these two phases, $\Gamma_R(t)$ and $\Gamma_I(t)$ can also be calculated 
from the correlation function~\cite{Mahan90,Leggett87,Makhlin01}
\begin{align}
\phantom{}_E\bra{0} e^{\Phi(t)}\, e^{\Phi^{\dag}(0)}\ket{0}_{E} 
\equiv e^{-\Gamma(t)}  \,,
\end{align}
where $\Phi\equiv -\frac{2\lambda}{\omega}(\ad - a)$ and 
$\Phi(t) = e^{i\omega t\,\ad a}\, \Phi\, e^{-i\omega t\,\ad a}$.
It is straightforward to obtain $\Gamma(t)$ given by 
\begin{subequations}
\begin{align}
\Gamma(t) &= \Gamma_R(t) + i \Gamma_I(t)  \\
          &= \left(\frac{2\lambda}{\omega}\right)^2
             \bigl[\, (1-\cos\omega t) - i\sin\omega t \,\bigr]\,.
\end{align}
\end{subequations}
The roles of the real and imaginary parts of $\Gamma(t)$ are explained below.
Then, in the limit of $\omega\gg\lambda$ Eq.~(\ref{Eq:psi_t}) can be written 
as a product state of the two qubits and the harmonic oscillator
\begin{align}
\ket{\Psi(t)} 
\approx \Bigl[ \,a\ket{00} + d\ket{11}
                + e^{-i2\theta t} (b\ket{01} + c\ket{10})\,
        \Bigr] \otimes \ket{0}_E\,.
\end{align}
The harmonic oscillator remains isolated always from the system, but it induces 
the indirect interaction between the two qubits and thus entanglement between 
them~\cite{cubitt03}. For a pure two qubits, concurrence, an entanglement measure, 
reads $C_{\rm ideal} = 2|ad - e^{i4\theta t}bc|$~\cite{Wootters98}. 
Here the subscript, ideal, stands for the case of the qubits in a pure state. 
If $a=b=c=d=1/2$, we have $C_{\rm ideal}=|\sin 2\theta t|$ as shown 
in Fig.~\ref{Fig2}.

The above analysis shows that the common harmonic oscillator with the high 
frequency, compared with the coupling constant to qubits, induces 
an inter-qubit coupling and can be used as a quantum data bus. However, 
the induced qubit-qubit coupling $I_{\rm eff}$ is very small, so that 
the operation time of two-qubit gates is much longer than that of 
single-qubit gates. Notice that some superconducting qubits use this 
kind of inter-qubit coupling scheme~\cite{Makhlin01}. As shown below, 
we find that a much faster two-qubit gate operation is possible and 
the maximally entangled state of the two qubits is obtainable even without 
the restriction of $\omega\gg\lambda$.

The harmonic oscillator plays two competing roles: (i) making two qubits 
entangled by inducing the two-qubit coupling $I_{\rm eff}$, (ii) having 
two qubits decoherent through entanglement with them, i.e. $\Gamma_R(t)$. 
The role of the imaginary part $\Gamma_I(t)$ is different from that of 
the real part $\Gamma_R(t)$. While $\Gamma_R(t)$ makes the two qubit decoherent, 
i.e. decaying the off-diagonal elements, $\Gamma_I(t)$ causes the two-qubit 
coupling to fluctuate as shown in Fig.~\ref{Fig2}(b). These could be 
uncovered by examining the reduced density matrix of the two qubits
\begin{align}
\label{Eq:density_matrix}
\rho(t)&= \frac{1}{4} 
\begin{bmatrix}
1 &e^{i2\theta t -\Gamma^*(t)} & e^{i2\theta t - \Gamma^*(t)} 
                               & e^{-4\Gamma_R(t)}                  \\[6pt]
e^{-i2\theta t -\Gamma(t)}     & 1 & 1 & e^{-i2\theta t -\Gamma(t)} \\[6pt]
e^{-i2\theta t -\Gamma(t)}     &1  & 1 & e^{-i2\theta t -\Gamma(t)} \\[6pt]
e^{-4\Gamma_R(t)}  &e^{i2\theta t -\Gamma^*(t)} 
                   &e^{i2\theta t -\Gamma^*(t)} &1
\end{bmatrix},
\end{align}
which is obtained by partial tracing over the environment,
$\rho(t)={\rm Tr}_E\bigl\{\ket{\Psi(t)}\bra{\Psi(t)}\bigr\}$.
Entanglement of the two qubits in a mixed state is measured by 
the concurrence $C =\max\{r_1 - r_2 - r_3 - r_4,0\}$ where $r_i^2$ 
are eigenvalues of $\rho\cdot\sigma_{1y}\otimes\sigma_{2y}\cdot \rho^*
\cdot\sigma_{1y}\otimes\sigma_{2y}$~\cite{Wootters98}. 
Because the total state is in a pure state, 
entanglement of the two qubits and the harmonic oscillator is quantified 
by the von Neumann entropy of the two qubits 
$S = -{\rm Tr}(\rho\log\rho)$.

As $\omega/\lambda$ becomes large, $C$ goes to $C_{\rm ideal}$ and $S$ to $0$ 
as depicted in Fig.~\ref{Fig2}. This is explained by means of 
the Born-Oppenheimer approximation that is based on the assumption of 
the weak coupling between the two qubits and the environment. 
The frequency $\omega$ of the harmonic 
oscillator is larger than that of the two qubits, $\theta$. 
The harmonic oscillator oscillates very fast in comparison with the two qubits. 
So the two qubits feel the harmonic oscillator stays in the same state. 
However, the condition of $\omega\gg\lambda$ is not a unique way to make 
the two qubits entangled maximally. Surprisingly, we find the maximum 
entanglement of the two qubits, $C =1 $, at $\theta t= \pi/4$ under 
the condition that $\omega/\lambda = 4\sqrt{n}$ with $n=1,2,\dots$ as shown 
in Figs.~\ref{Fig2}(a) and \ref{Fig2}(d). This is due to the fact that 
the frequency $\theta$ for entangling the two qubits is commensurate 
with the frequency $\omega$ of the harmonic oscillator. 
If the condition of $\omega/\lambda = 4\sqrt{n}, n=1,2,\dots$ 
is not met, then the concurrence $C$ does not reach 1 
at $\theta t=\pi/4$ and 0 at 
$\theta t=\pi/2$ due to $\Gamma(t)$ as shown in Fig.~\ref{Fig2}(b). 
The oscillation period of $C$ (red solid line) does not coincide with 
that of the ideal case $C_{\rm ideal}$ (thick black line). 
This implies that the two qubit coupling fluctuates due to $\Gamma_I(t)$. 
For $\omega/\lambda < 1$, our numerical study shows that the two qubits could 
not be entangled. Fig.~\ref{Fig2}(c) shows the case of $\omega=\lambda$.  

Fig.~\ref{Fig2}(d) shows two competing roles of the harmonic oscillator. 
Let us define the average concurrence $C_{\rm avg}$ for a period $\tau\equiv 
\pi/2\theta$ by $C_{\rm avg} \equiv \frac{1}{\tau}\int_0^{\tau} C(t)\,dt$.
For the first period $0\le \theta t <\pi/2$, the maximum concurrence 
$C_{\rm max}$ reaches the peak at positive integers $n$ and the average 
concurrence $C_{\rm avg}$ increases as ${\omega}/{\lambda}$ increases. 
In contrast, the maximum entropy $S_{\rm max}$ and the average entropy 
$S_{\rm avg} \equiv \frac{1}{\tau}\int_0^\tau S(t)\,dt$ decrease. 
Due to the commensuration of two frequencies 
$\omega$ and $\theta$, $S_{\rm avg}$ and $C_{\rm avg}$ show the behavior 
of the stair case. 

\paragraph*{Two-spin boson model.---} In the single mode case, the dynamics 
is reversible. If the single harmonic oscillator is replaced by the
bosonic bath, the dynamics of qubits becomes irreversible in the sense that 
a pure initial qubit state becomes mixed and never returns to a pure state. 
The total Hamiltonian of the two-spin boson model is given by 
\begin{align}
\label{Eq:two_spin_boson}
H = \sum_{j=1}^{N} \hbar\omega_j a_j^\dag a_j
  + (\sigma_{1z} + \sigma_{2z})\,
    \sum_{j=1}^{N} \hbar\lambda_j (a_j^\dag + a_j) \,,
\end{align}
where the first term describes the bosonic bath and the second the interaction 
between the two qubits and the bosonic bath (environment). Here it is assumed 
that $N$ goes to infinity.  The Hamiltonian of two qubits 
$H_q = \sum_{i=1,2}\left(B_{ix}\sigma_{ix} + B_{iz}\sigma_{iz}\right)$  
is set to be zero by turning off an external magnetic field. Also it is assumed 
that no direct two qubit coupling is available.

\begin{figure}
\includegraphics[scale=0.9]{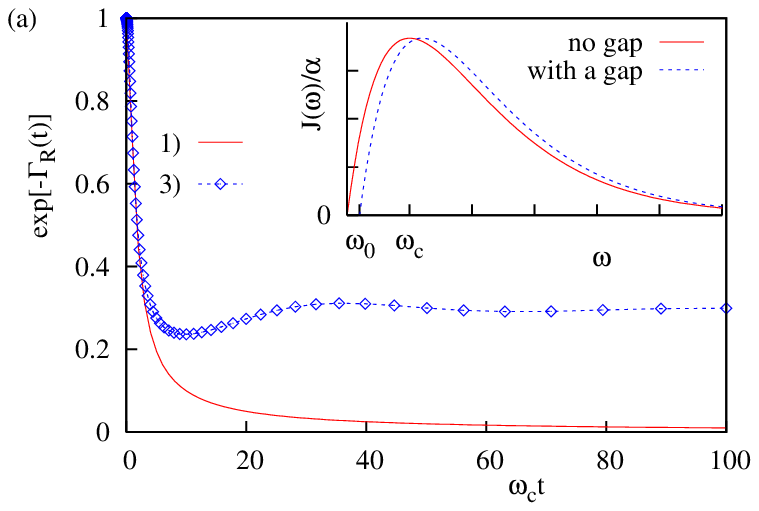}\\
\includegraphics[scale=0.9]{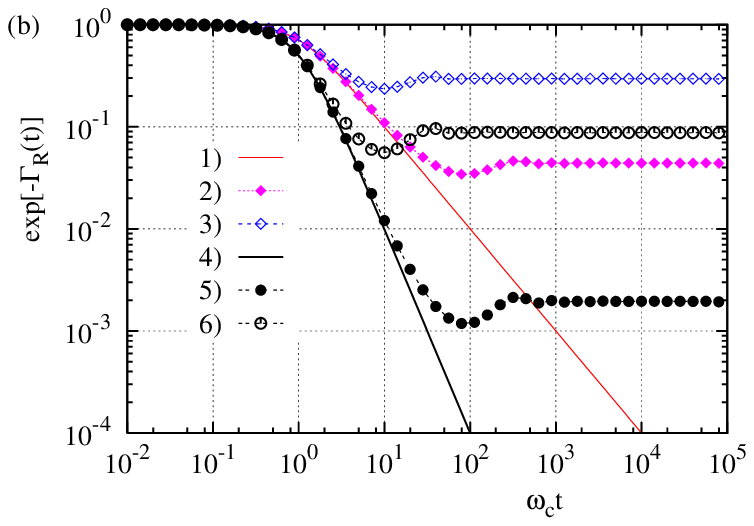}\\
\includegraphics[scale=0.9]{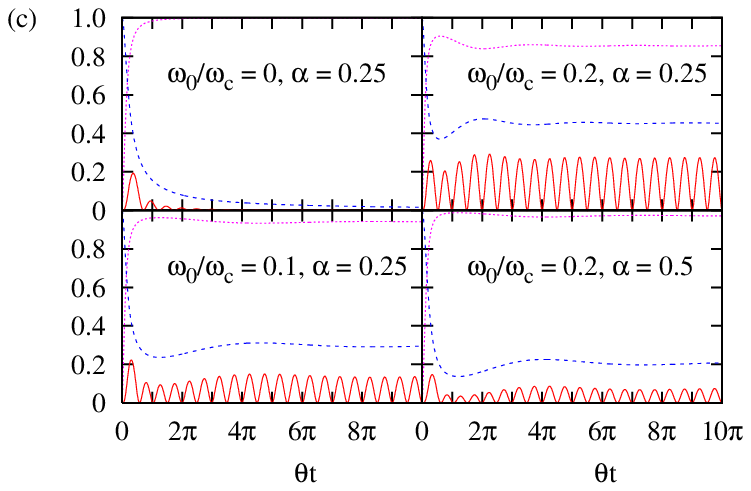}
\caption{(color online). (a) $\exp[-\Gamma_R(t)]$ as a function of $\omega_c t$
         and (b) its log scale plot. Legends $1),\dots,6)$ refer to 
         $(\omega_0/\omega_c,\alpha)$ = (0,0.25), (0.01,0.25), (0.1,0.25), 
         (0,0.5), (0.01,0.5), (0.1,0.5), respectively. Inset of (a): 
         the spectral density function $J(\omega)$ with a gap $\omega_0$. 
         (c) Concurrence $C$ (solid), entropy $2S/3$ (dotted), and 
         $\exp[-\Gamma_R(t)]$ (dashed) as a function of $\theta t$.}
\label{Fig3}
\end{figure}

The environment is characterized by the spectral density function
$J(\omega) = \sum_j\lambda_j^2 \,\delta(\omega - \omega_j)$
~\cite{Leggett87,Weiss98}.
As shown in the inset of Fig.~\ref{Fig3}-(a), let us consider an Ohmic 
environment with a gap $\omega_0$ 
and an exponential cutoff function of the cutoff frequency $\omega_c$
\begin{align}
J(\omega) 
= \alpha(\omega-\omega_0)\,e^{-(\omega-\omega_0)/\omega_c}\,
  \theta(\omega-\omega_0)\,,
\label{Eq:spectral_density}
\end{align}
where $\alpha$ is the coupling parameter and 
$\theta(x)$ a step function.

Along the same lines of the single mode case, the canonical transformation
\begin{align}
e^X = \exp \left[ (\sigma_{z1} + \sigma_{z2})
                 \sum_j\frac{\lambda_j}{\omega_j}(\ad_j -a_j) 
           \right]\,
\end{align}
eliminates the coupling between the two qubits and the common environment 
in the Hamiltonian~(\ref{Eq:two_spin_boson}) and induces the indirect coupling 
between the two qubits. The transformed Hamiltonian becomes
\begin{align}
\label{Eq:eff_Hamil}
\widetilde{H} 
 = \sum_j\hbar\omega_j\ad_ja_j 
  - I_{\rm eff}\sigma_{1z}\sigma_{2z} - I_{\rm eff}\,,
\end{align}
where the induced coupling is given by 
\begin{align}
\label{Eq:I_eff}
I_{\rm eff} &\equiv \hbar\sum_j\frac{2\lambda^2_j}{\omega_j} 
            = 2\hbar\int_0^\infty \frac{J(\omega)}{\omega}\, d\omega \,.
\end{align}
For the Ohmic environment without a gap, $\omega_0 =0$, 
we have $I_{\rm eff} = 2\hbar\alpha\omega_c$.

Let us investigate the dynamics of entanglement. For an initial state 
$\ket{\Psi(0)} = \ket{\psi(0)}\otimes \ket{\mathbf{0}}_E$
where $\ket{\mathbf{0}}_E=\ket{0_1,0_2,\dots,0_N}$ is the ground state of 
$N$ harmonic oscillators, the state at time $t$ is given by
\begin{align}
\ket{\Psi(t)} 
 & = e^{i\theta t - i\Gamma_I(t)}\,
    \Bigl[\, a\ket{00}\otimes\ket{{\bm{\alpha}}(t)}_E
   + d\ket{11}\otimes\ket{-{\bm{\alpha}}(t)}_E \nonumber\\
 &+ e^{-i2\theta t+i\Gamma_I(t)} (\,b\ket{01} + c\ket{10}\,) 
    \otimes\ket{\bm{0}}_E\, \Bigr] \,,
\label{Eq:total_state}
\end{align}
where $\ket{\bm{\alpha}(t)}_E 
\equiv \ket{\{\textstyle{\frac{\lambda_i}{\omega_i}(e^{-i\omega_it} -1)\}}}$ 
is the collection of the coherent states.  The imaginary part of $\Gamma(t)$ 
becomes
\begin{subequations}
\begin{align}
\Gamma_I(t) &= \sum_j \left(\frac{2\lambda_j}{\omega_j}\right)^2\sin\omega_jt \\
            &= 4\int_0^{\infty}\frac{J(\omega)}{\omega^2}\sin(\omega t)
            \,d\omega\,.
\end{align}
\label{Eq:Gamma_i}
\end{subequations}
Also we obtain the overlap 
$\phantom{}_E\braket{\bm 0}{{\bm\alpha}(t)}_E \equiv \exp[-\Gamma_R(t)]$ where
\begin{align}
\Gamma_R(t)
= 4\int_0^{\infty}\frac{J(\omega)}{\omega^2}(1-\cos\omega t) 
            \,d\omega\,.
\label{Eq:gamma}
\end{align}
From Eq.~(\ref{Eq:total_state}), the density matrix of the two qubits becomes 
the same form of Eq.~(\ref{Eq:density_matrix}).  
For the Ohmic spectral density~(\ref{Eq:spectral_density}) without a gap, 
$\omega_0=0$, we have~\cite{Leggett87,Gradshteyn79}
\begin{subequations}
\begin{align}
\Gamma_R(t) &= 2\alpha\ln[1+(\omega_ct)^2] \,, \\
\Gamma_I(t) &= 4\alpha\tan^{-1}(\omega_c t) \,.
\end{align}
\end{subequations}
The slopes of solid lines in Fig.~\ref{Fig3}(b) are $4\alpha$.  
In this case $\exp[-\Gamma_R(t)]$ goes to zero as time goes on, 
but never returns to 1 in contrast to the single-mode case.
Then the density matrix of the two qubits eventually becomes a mixed state 
with the maximum entropy $S=2/3$. For the environment without a gap, 
the two qubits become entangled and disentangled only for the transient times 
not in the ``steady state" as shown in Fig.~\ref{Fig3}(c).
Here the term ``steady state" is used in the sense that in the steady state 
the entropy $S(t)$ of the two qubits does not change in time~\cite{Breuer02}. 
In this case, there is no further information flow from the two qubit to 
the environment.

It is clear that in the steady state the two qubits could be entangled and 
disentangled only if $\exp[-\Gamma_R(t)]$ remains finite. How is it possible? 
We find that the small gap $\omega_0 \ll \omega_c$ of the Ohmic environment 
makes $\exp[-\Gamma_R(t)]$ finite in the steady state and the two qubits 
entangled.  Fig.~\ref{Fig3} shows that for the Ohmic environment with a gap 
the decoherence term $\exp[-\Gamma_R(t)]$ does not vanish in the steady state. 
The role of the gap is to prevent the complete drain of information from 
the qubits to the bath. We would like to mention a similar phenomenon, that is,
the inhibition of the spontaneous emission of an atom in photonic-band-gap 
materials~\cite{John94}. Notice that the gap of the spectral density function 
causes the phenomenon called false decoherence~\cite{Leggett04}. 
It would be interesting to derive a master equation of the two qubits in 
the Lindblad form when the spectral density of the environment has a gap. 

\begin{figure}
\includegraphics[scale=0.85,angle=0]{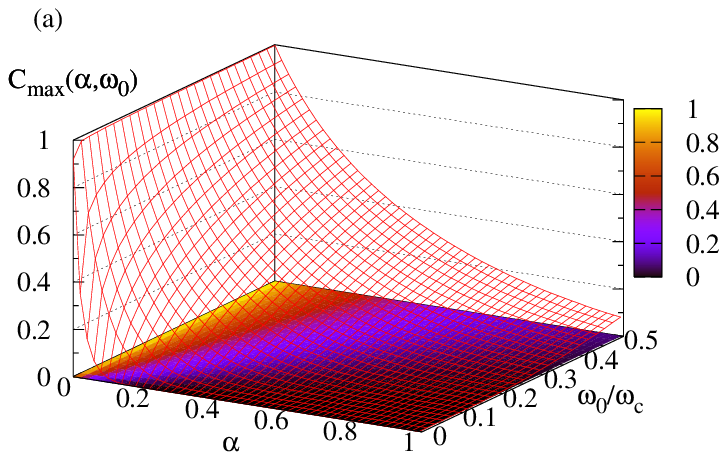}\\[-20pt]
\includegraphics[scale=0.85,angle=0]{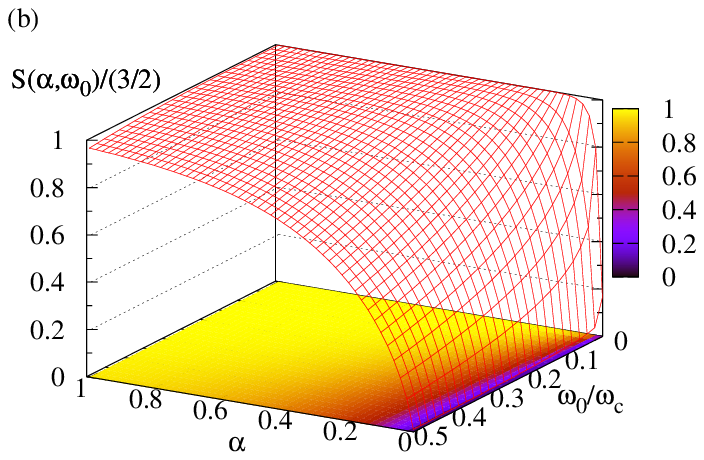}\\[-20pt]
\includegraphics[scale=0.85,angle=0]{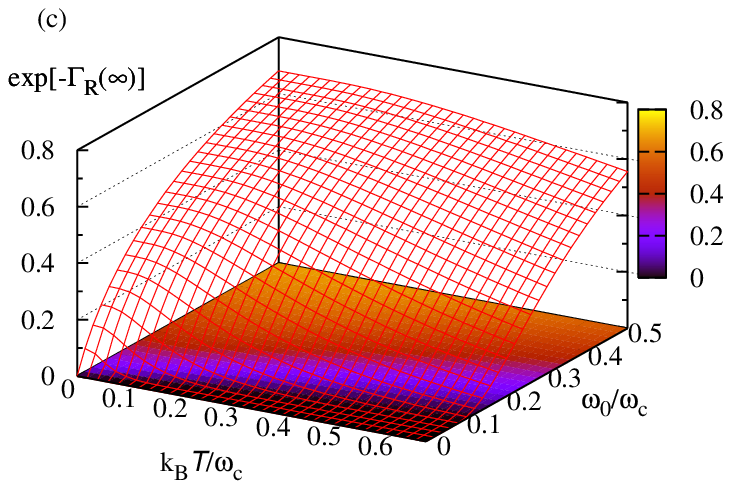}
\caption{(color online). (a) Maximum concurrence of the two qubits 
(b) entropy of two qubits at equilibrium as a function of $\alpha$ and 
$\omega_0/\omega_c$. (c) $\exp[-\Gamma_R(\infty)]$ as a function of temperature 
$T$ and $\omega_0/\omega_c$ for $\alpha =0.25$.}
\label{Fig4}
\end{figure}

Figs.~\ref{Fig4}(a) and \ref{Fig4}(b) show the maximum concurrence $C_{\rm max}$
and the entropy $S$ in the steady state at zero temperature. 
$C_{\rm max}$ ($S$) decreases (increases) as $\alpha$ increases even 
for $\omega_0 \ne 0$. The effect of the non zero 
temperature of the environment is shown in Fig.~\ref{Fig4}(c). 
For finite temperatures, we take the initial density matrix $W(0)$ of the 
total system given by
\begin{align}
W(0) = \ket{\psi(0)}\bra{\psi(0)}\otimes \rho_E\,,\quad
\rho_E = \frac{e^{-H_E/k_BT}}{Z_E}\,,
\end{align}
where the environment is in a thermal equilibrium state $\rho_E$ at 
temperature $T$.
Here $H_E = \sum_i\hbar\omega_i\ad_i a_i$ and $Z_E$ is the partition function
of the environment. At finite temperatures, the decoherence factor is given 
by~\cite{Leggett87,Mahan90,Makhlin01}
\begin{align}
\Gamma_R(t) = 4\int_0^{\infty}\frac{J(\omega)}{\omega^2}
\coth(\textstyle{\frac{\hbar\omega}{2k_BT}}) (1-\cos\omega t)\,
d\omega
\end{align}
which decreases the effect of the gap as shown in Fig.~\ref{Fig4}(c).
However, the real part $\Gamma_R(t)$ is identical to Eq.~(\ref{Eq:Gamma_i}).

We would like to make some remarks. First, it is easy to obtain the induced 
interaction for many qubits interacting with a common environment. 
It has the form of pair-wise interactions $\widetilde{H}\propto 
I_{\rm eff}\sum_{ij}\sigma_{iz}\sigma_{jz}$. In this paper
no external magnetic fields are applied on qubits, i.e., ${\bf B}_{i}$, $i=1,2$.
For ${\bf B}_i\ne 0$, one can also obtain the effective Hamiltonian
using the flow equation renormalization group~\cite{Oh04}, which is 
a generalized canonical transformation.

Second, there is an interesting physical systems similar to the two-spin 
boson model. Ref.~\cite{Craig04} considers two spins of the two quantum 
dots coupled with a 2-dimensional electron gas. In this case the induced 
interaction is nothing but the Ruderman-Kittel-Kasuya-Yosida (RKKY) 
interaction. However, the spin on a quantum dot becomes entangled with 
a 2-dimensional electron gas~\cite{Oh05}.  Also the density of states of 
an electron gas has no gap.  According to our theory in this paper, 
although the two spins are 
coupled via the RKKY interaction, their state is not pure but mixed. 
A further study should be done in order to know whether the RKKY interaction 
could be used for a two-qubit coupling scheme in solid state quantum 
computers~\cite{Craig04}.

In conclusion, we have studied the two qubits interacting with a common 
environment with a gap, described by a two-spin-boson model. 
For the environment of a single harmonic oscillator, 
if the induced two-qubit coupling strength is commensurate 
with the frequency of the harmonic oscillator, the two qubits could 
become maximally entangled. Also we discussed the roles of the real and
imaginary parts of $\Gamma(t)$.  In the case of the two-spin boson model 
with the spectral density function of the gap $\omega_0$, at the steady state 
the two qubits become entangled and disentangled due to the indirect 
interaction between qubits induced by the environment. 
Our result shows explicitly that two qubits can be entangled due to 
the environment with a gap while the environment itself is not entangled 
with qubits. 

\acknowledgments

We would like to thank Howard Lee and Daniel Braun for helpful comments.
This research was supported by a Grant(TRQCQ) from the Ministry of Science 
and Technology of Korea and by Korea Research Foundation Grant KRF-2004-041-C0089. 
This work was also partially supported by Korea Research Foundation 
Grant KRF-2002-070-C00029.

\end{document}